\documentclass[twoside,12pt]{article}
\usepackage{epsfig}

\def\Journal#1#2#3#4{{#1} {#2} (#4) #3 }

\def\NPA{{\em Nucl. Phys.} A}

\def\PLB{{\em Phys. Lett.} B}

\def\PRL{\em Phys. Rev. Lett.}

\def\PREP{\em Phys. Rep.}

\def\PRC{{\em Phys. Rev.} C}

\newcommand{\be}{\begin{equation}}
\newcommand{\ee}{\end{equation}}
\newcommand{\bea}{\begin{eqnarray}}
\newcommand{\eea}{\end{eqnarray}}

\begin{document}

\title{ \vspace{1cm} Probing the Symmetry Energy with Heavy Ions}

\author{W.G.\ Lynch,$^{1,2}$ M.B.\ Tsang,$^{1,2}$ Y.\ Zhang,$^{1,3}$ P.\
Danielewicz,$^{1,2}$\\
M.\ Famiano,$^4$ Z.\ Li,$^3$ and A. W.\ Steiner,$^{1,2}$\\
\\
$^1$JINA and NSCL, Michigan State University,\\
$^2$Dept. of Physics and Astronomy, Michigan State University,\\
$^3$China Institute of Atomic Energy, Beijing,\\
$^4$Physics Department, Western Michigan University.}

\maketitle
\begin{abstract} Constraints on the EoS for symmetric matter (equal neutron and proton numbers) at supra-saturation densities have been extracted from energetic collisions of heavy ions. Collisions of neutron-deficient and neutron-rich heavy ions now provide initial constraints on the EoS of neutron-rich matter at sub-saturation densities. Comparisons are made to other available constraints.
\end{abstract}
\section{Introduction}
The nuclear EoS is a fundamental property of nuclear matter that describes the relationships between the energy, pressure, temperature, density and isospin asymmetry  $\delta\equiv(\rho_{n}-\rho_{p})/(\rho_{n}+\rho_{p})$
 for a nuclear system. (Here, $\rho_{n}$ and $\rho_{p}$ are the neutron and proton densities, respectively.) It can be divided into a symmetric matter contribution that is independent of the isospin asymmetry and a symmetry energy term, proportional to the square of the asymmetry, that describes the entire dependence of the EoS on asymmetry. Investigations that provide an improved understanding of this term will also provide an improved understanding of masses \cite{Dan03,Dan08,Ste05}, fission barriers, energies of isovector collective vibrations \cite{Dan03,Kli07}, and the thickness of the neutron skins of neutron-rich nuclei \cite{Bro00,Hor01}.

Macroscopic quantities of asymmetric nuclear matter exist over a wide range of densities in neutron stars and in type II supernovae \cite{Lat04,Lat01}.  Experimental information about the EoS can help to provide improved predictions for neutron star observables such as stellar radii and moments of inertia, crustal vibration frequencies \cite{Lat04,Vil04}, and neutron star cooling rates \cite{Ste05,Lat04} that are currently being investigated with ground-based and satellite observatories.  For many of these observables, the absence of strong constraints on the symmetry energy term of the EoS engenders major theoretical uncertainties. Consequently, the goal of determining the EoS has been a major motivation for many X-ray observations of neutron stars \cite{Ixo08}. Experimental investigations are needed to check whether constraints derived from future neutron star observations can be supported by laboratory measurements.

The total energy per nucleon (i.e. the Equation of State (EoS)) of cold nuclear matter can be written as the sum of a symmetry energy term and the energy per nucleon of symmetric matter,

\be
E(\rho,\delta)=E_{0}(\rho)+E_{\delta}; E_{\delta}=S(\rho)\cdot\delta^{2} \; , \label{eq:EOS}
\ee

\noindent where $S(\rho)$ describes the density dependence of the symmetry energy term. Measurements of isoscalar collective vibrations, collective flow and kaon production in energetic nucleus-nucleus collisions have constrained the equation of state for symmetric matter, $E_{0}(\rho)$, and the pressure,  $P=\rho^{2}\cdot(\partial E_{0}(\rho)/\partial\rho|_{s})$, at $T=0$ and densities ranging from saturation density to five times saturation density \cite{Dan02,Fuc06,You97}.

\begin{figure}[tb]
\begin{center}
\begin{minipage}[t]{10.5 cm}
\epsfig{file=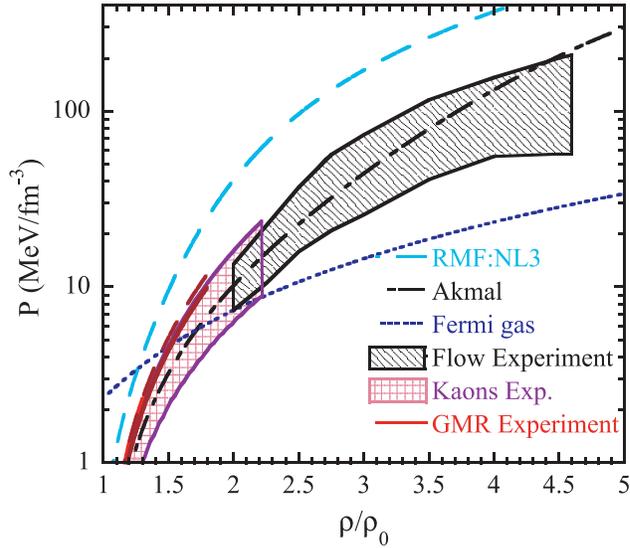,scale=0.6}
\end{minipage}
\begin{minipage}[t]{16.5 cm}
\caption{Constraints on the symmetric matter EoS from heavy ion collisions. The experimental constraints (cross-hatched and shaded regions) and the theoretical calculations are discussed in the text. \label{fig1}}
\end{minipage}
\end{center}
\end{figure}

Fig.~\ref{fig1} illustrates some of these constraints graphically. The shaded region extending from $\rho/\rho_{0}=2$ to $\rho/\rho_{0}=4.5$ shows the range of pressures in cold symmetric nuclear matter for EoS's that describe the available transverse and elliptical flow data \cite{Dan02}. The center of the shaded region extending from $\rho/\rho_{0}=1.2$ to $\rho/\rho_{0}=2.2$ shows the pressure - density relationship that best describes the available kaon production data \cite{Fuc06}. The range of densities for the kaon constraint represents an educated guess based on available calculations \cite{Fuc06}; systematic studies to establish this range of pressures have not been performed. The dashed line extending approximately from $\rho/\rho_{0}=1.2$ to $\rho/\rho_{0}=1.7$ extrapolates the pressures   consistent with the Giant Monopole Resonance data to higher densities. Various theoretical curves labeled RMF:NL3 \cite{Lal97}, Akmal \cite{Akm98}, and Fermi Gas illustrate the larger range of theoretical predictions for the EoS at supra-saturation densities. Experimental constraints can discriminate between predicted behaviors of the EoS at supra-saturation densities and show which are more likely.

The extrapolation of the EoS to neutron-rich matter depends on $S(\rho)$, which has comparatively few experimental constraints \cite{Bro00}. Many recent efforts to constrain the density dependence of the symmetry energy have focused on its behavior near saturation density. There, one may expand the symmetry energy, $S(\rho)$, about the saturation density as follows,
\be
S(\rho)=S_{0}+\frac{L}{3}(\frac{\rho-\rho_{0}}{\rho_{0}})+\frac{K_{sym}}{18}(\frac{\rho-\rho_{0}}{\rho_{0}})^{2}+...
 \; , \label{eq:expansion}
\ee

\noindent where $L$ and $K_{sym}$ are slope and curvature parameters at $\rho_{0}$.  The slope parameter, $L$, is related to $p_{0}$, the pressure from the symmetry energy for pure neutron matter at saturation density via $ L=3\rho_{0}|dS_{\rho}/d\rho|_{\rho_{0}}=[3/\rho_{0}]/p_{0}$. The symmetry pressure, $p_{0}$, provides the baryonic contribution to the pressure in neutron stars at saturation density \cite{Ste05}, where the energy of symmetric matter, $E_{0}(\rho)$, contributes no pressure, and it is also related to the neutron skin thickness, $\delta R_{np}$, of neutron rich heavy nuclei including $^{208}Pb$ \cite{Bro00,Hor01}.

\section{Experimental Probes of $S(\rho)$}

Both symmetry and total binding energy terms in the nuclear semi-empirical binding energy formulae reflect averages of $E(\rho,\delta)$  over the densities of nuclei \cite{Dan03}. The values of surface and volume symmetry energy terms obtained over fits of such formula to measured masses provide some sensitivity to the density dependence of $S(\rho)$ near saturation density \cite{Dan03,Dan08}. Giant resonances, low-lying electric dipole excitations and the difference between neutron and proton matter radii may also provide sensitivity to the density dependence of $S(\rho)$ near saturation density \cite{Kli07,Hor01,Tli07,Pie04}.

Nuclear collisions can transiently produce both sub-saturation and supra-saturation
density variations. The symmetry energy has been recently probed at sub-saturation densities in collisions via measurements of isospin diffusion \cite{Tsa04,Liu07}, and of double ratios involving neutron and proton energy spectra \cite{Fam06}. Future comparisons of the spectra and flows of neutrons vs. protons and negative vs. positive pions can provide constraints on the symmetry energy at supra-saturation densities \cite{Bal08,Bar05}.  Such observables reflect the transport of nucleons under the influence of nuclear mean fields and of the collisions induced by the residual interactions, both of which can be described by transport theory. Focusing on sub-saturation densities, Chen et al., obtained a reasonable description of isospin diffusion data with a symmetry energies of approximate form with  =0.69-1.1 \cite{Bal08,Che05} using a Boltzmann-Uehling-Uhlenbeck (BUU) transport model IBUU04. While the double ratios of neutron and proton energy spectra \cite{Fam06} can also provide constraints at sub-saturation densities, these data have not been reproduced by the IBUU04 model \cite{Bal08}.

\section{Constraints on $S(\rho)$ }

Recently, we have explored how isospin diffusion and n-p double ratios depend on both $S_{0}$ and $L$ parameters of $S(\rho)$ using the Improved Quantum Molecular Dynamics (ImQMD) transport model\cite{Tsa08}. Three observables, the neutron/proton double ratio \cite{Fam06}, the isospin transport ratio at beam rapidity obtained from the isoscaling parameters \cite{Tsa04}, and the isospin transport ratios at various rapidities obtained from $^{7}Li/^{7}Be$ mirror nuclei ratios \cite{Liu07} were compared to ImQMD calculations for different assumptions about the density dependence of the symmetry energy \cite{Tsa08}.

\begin{figure}[tb]
\begin{center}
\begin{minipage}[t]{10.5 cm}
\epsfig{file=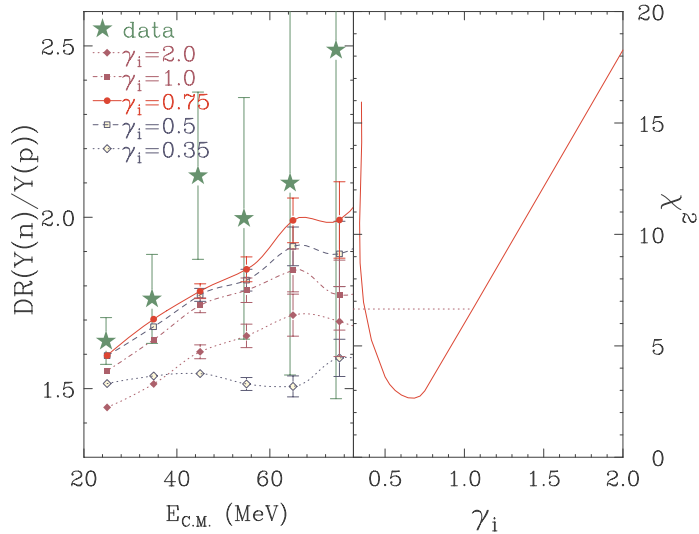,scale=0.6}
\end{minipage}
\begin{minipage}[t]{16.5 cm}
\caption{Left panel: Comparison of experimental double neutron-proton ratios (star symbols), as a function of nucleon center of mass energy, to ImQMD calculations (lines) with different density dependencies of the symmetry energy, parameterized by $\gamma_{i}$ in Eq. (1). Right panel: A plot of $\chi^{2}$ as a function of  $\gamma_{i}$. \label{fig2}}
\end{minipage}
\end{center}
\end{figure}

Detailed description of the model and its application to the double ratio data can be found in ref. \cite{Zha08}. For brevity, we limit our discussion here to the parameterization of the symmetry energy used in our calculations, which is of the form
\be
S(\rho)=\frac{C_{c,k}}{2}(\frac{\rho}{\rho_{0}})^{2/3}+\frac{C_{c,p}}{2}(\frac{\rho}{\rho_{0}})^{\gamma_{i}}
 \; , \label{eq:qmdeos}
\ee

\noindent Unless indicated otherwise, the kinetic and potential parameters are $C_{s,k}=25 MeV$, $C_{s,p}=35.2 MeV$ and the symmetry energy at saturation density, $S_{0} =S(\rho_{0}) =30.1 MeV$.

We first turn our attention to the interpretation of neutron/proton double ratio data within the ImQMD model \cite{Zha08}. This observable derives its sensitivity to the symmetry energy from the opposite sign of the symmetry potential for neutrons as compared to protons \cite{Bal08}. First experimental comparisons of neutron to proton spectra in ref. \cite{Fam06} used a double ratio in order to reduce sensitivity to uncertainties in the neutron detection efficiencies and sensitivity to relative uncertainties in energy calibrations of neutrons and protons. This double ratio,
\be
DR(Y(n)/Y(p))=R_{n/p}(1)/R_{n/p}(2)=\frac{dM_{n}(1)/dE_{c.m.}}{dM_{p}(1)/dE_{c.m.}}\cdot\frac{dM_{p}(2)/dE_{c.m.}}{dM_{n}(2)/dE_{c.m.}} \; , \label{eq:DR}
\ee
is constructed from the ratios of energy spectra, $dM_{n}/dE_{c.m.}$ dM/dEC.M, of  neutrons and protons for two systems 1 $\equiv$ $^{124}Sn$ + $^{124}Sn$ and 2 $\equiv$ $^{112}Sn$ + $^{112}Sn$ with different isospin asymmetries. The stars in the left panel of Fig.~\ref{fig2} show the resulting neutron-proton double ratios measured at $70^{o}\leq\Theta_{CM}\leq110^{o}$ as a function of the center-of-mass (C.M.) energy of nucleons emitted from central collisions \cite{Fam06}.

We have performed calculations for these two systems. Within statistical uncertainties, the double ratio observable, $DR(Y(n)/Y(p))$, is nearly independent of  impact parameter for $1\leq   b\leq5 fm$. The lines in the left panel of Fig.~\ref{fig2} show double ratios, averaged over $b=1, 2, 3 fm$, vs. the C.M. energy of nucleons for $\gamma_{i}=0.35, 0.5, 0.75, 1$ and $2$. The theoretical uncertainties in Fig.~\ref{fig2} are statistical. Despite the large experimental uncertainties for higher energy data, these comparisons rule out both very soft ($\gamma_{i}=0.35$, dotted line with closed diamond points) and very stiff ($\gamma_{i}=2$, dotted line with open diamond symbols) density-dependent symmetry terms. The right panel shows the $\gamma_{i}$ dependence of the total $\chi^{2}$ computed from the difference between predicted and measured double ratios. We determine, to within a $2\sigma$ uncertainty, parameter values of $0.4\leq\gamma_{i}\leq1.05$ corresponding to an increase in $\chi^{2}$ by 4 above its minimum near $\gamma_{i} = 0.7$.

Decreasing $\gamma_{i}$ corresponds to higher symmetry energy at sub-saturation densities, which enhances the emission of neutrons. Similarly, decreasing  $\gamma_{i}$ corresponds to increasing $DR(Y(n)/Y(p))$. In the limit of very small  $\gamma_{i} << 0.35$, however, the system completely disintegrates and $DR(Y(n)/Y(p))$ decreases towards the free nucleon value $N_{total}/Z_{total} =1.2$. As a consequence of these two competing effects, the double ratio attains maximum values at about  $\gamma_{i}\sim 0.7$.

The density dependence of the symmetry energy has also been probed in peripheral collisions between two nuclei with different isospin asymmetries by examining the diffusion of neutrons and protons between them. Unless the two nuclei separate, this "isospin diffusion" would continue until the chemical potentials for neutrons and protons in both nuclei become equal. To isolate diffusion effects from pre-equilibrium emission, Coulomb effects and secondary decays, measurements of isospin diffusion compare "mixed" collisions, involving a neutron-rich nucleus $A$ and a neutron-deficient nucleus $B$, to the "symmetric" collisions involving $A+A$ and $B+B$. The degree of isospin equilibration in such collisions can be quantified by rescaling the isospin observable $X$ according to the isospin transport ratio $R_{i}(X)$ \cite{Tsa04}
 \be
R_{i}(X)=2\cdot\frac{X-(X_{A+A}+X_{B+B})/2}{X_{A+A}-X_{B+B}} \; , \label{eq:ri}
\ee
where $X$ is the isospin observable. Adopting the order of the notation, $A+B$ to designate $A$ and $B$ as the projectile and target, respectively, one expects $R_{i}(A+B)=R_{i}(A+A)$ and $R_{i}(B+A)=R_{i}(B+B)$ for $R$ values extracted from emission near projectile rapidities in the absence of isospin diffusion. In the opposite extreme, $R_{i}(A+B)=R_{i}(B+A)\approx0$ if isospin equilibrium is achieved.

\begin{figure}[tb]
\begin{center}
\begin{minipage}[t]{10.5 cm}
\epsfig{file=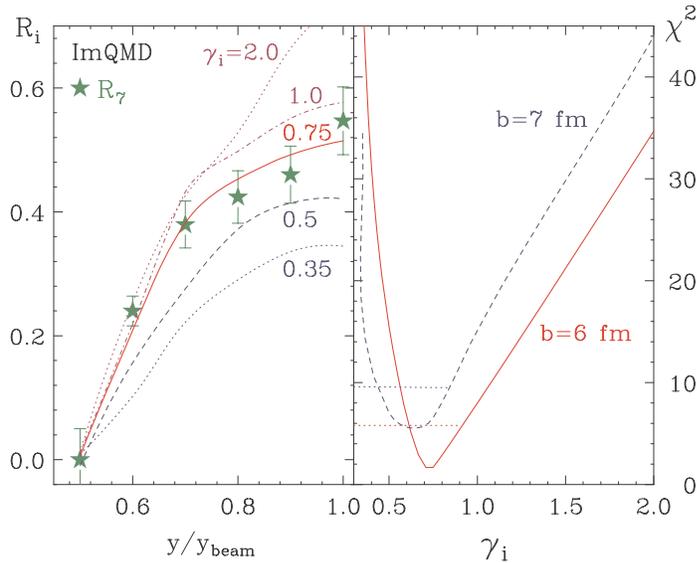,scale=0.6}
\end{minipage}
\begin{minipage}[t]{16.5 cm}
\caption{Left panel: Comparison of experimental isospin transport ratios obtained from the yield ratios of $A=7$ isotopes (star symbols), as a function of the rapidity, ImQMD calculations (lines) for $b=6 fm$. Right panel: $\chi^{2}$ analysis for $b=6 fm$ (solid curve) and $b=7 fm$ (dashed curve) as a function of $\gamma_{i}$. \label{fig4}}
\end{minipage}
\end{center}
\end{figure}

Eq. \ref{eq:ri} dictates that different observables, $X$, provide the same results if they are linearly related \cite{Liu07}. Experimental isospin transport ratios obtained with isoscaling parameters, $\alpha$, of charged particles of $Z=3-8$, $R_{i}(\alpha)$ and yield ratios of $A=7$ mirror nuclei, $R_{7} = R_{i}(X_{7}=ln(Y(^{7}Li)/Y(^{7}Be)))$ are consistent, ie. $R_{i}(\alpha)\cong R_{7}$, reflecting linear relationships between $\alpha$, $X_{7}$, and the asymmetry $\delta$ of the emitting source \cite{Liu07}. For emission at a specific rapidity y, we assume $R(\alpha)=R(\delta)=R_{7}$ to be valid, as has been confirmed experimentally \cite{Liu07} and theoretically for all statistical and dynamical calculations \cite{Tsa01,Bot02,Ono03}.  In the following, we calculate $\delta$ from the asymmetry of the fragments and free nucleons emitted at the relevant rapidity, but our conclusions do not significantly change if fragments alone are used to calculate $\delta$. For reference, we note that BUU calculations use the asymmetry of the projectile residues \cite{Tsa04} to construct $R_{i}(\delta)$.

Experimental isospin diffusion transport ratios, $R_{7}$, plotted as stars as a function of rapidity in the left panel of Fig.~\ref{fig4}, have been obtained using Eq. \ref{eq:ri} and the yield ratios of $^{7}Li$ and $^{7}Be$, \cite{Liu07}. It is estimated that the measurements include impact parameters ranging from $5.8$ to $7.5$ fm \cite{Liu07}. We have performed ImQMD calculations at impact parameters of $b=5, 6, 7,$ and $8 fm$. The lines in the left panel of Fig.~\ref{fig4} show ImQMD calculations at $b=6 fm$ . The corresponding $\chi^{2}$ analysis at $b=6$ (solid curve) and $7$ (dashed curve) $fm$ in the right panel displays sharp minima. Using the $\chi^{2}$  criterion adopted previously, the analysis favors the region $0.45\leq \gamma_{i}\leq 0.95$. A similar analysis of the experimental and calculated isospin diffusion transport ratios, $R_{i}(\alpha)$, has also been performed and favors the region $0.40\leq \gamma_{i}\leq 1.0$ see ref. \cite{Tsa08}.

\begin{figure}[tb]
\begin{center}
\begin{minipage}[t]{10.5 cm}
\epsfig{file=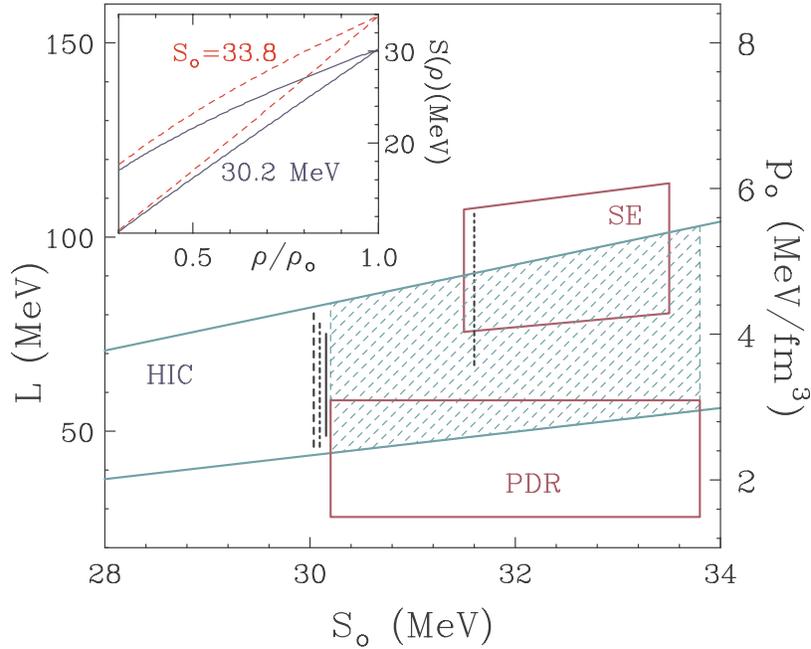,scale=0.7}
\end{minipage}
\begin{minipage}[t]{16.5 cm}
\caption{Representation of the constraints on parameters $S_{0}$ and $L$. The various lines represent constraints discussed in the text. Corresponding values of $p_{0}$ are given in the right axis. \label{fig5}}
\end{minipage}
\end{center}
\end{figure}

These constraints on the exponent $\gamma_{i}$ depend on the symmetry energy at saturation density, $S_{0}=  S(\rho_{0})$, as well as $L$ and $K_{sym}$. Within families of parameterizations for $S(\rho)$, $K_{sym}$ is largely determined by $L$. The present observables are more sensitive to $L$ than to $K_{sym}$; $L$ can therefore be determined more reliably than $K_{sym}$. We have preformed a series of ImQMD calculations at $b=6 fm$ with different values of $\gamma_{i}$ and $S_{0}$ to locate the approximate boundaries in the $S_{0}$ and $L$ plane that satisfy the $2\sigma$ criterion in the $\chi^{2}$ analysis of the isospin diffusion data. The two diagonal lines in Fig.~\ref{fig5} represents estimates in such effort. The sensitivity of constraints on $S_{0}$ and $L$ to differences between $m_{n}^{*}$ and $m_{p}^{*}$, and to the in-medium cross-sections have not been fully explored; we note that BUU calculations for isospin diffusion show much more sensitivity to $S_{0}$ and $L$ that to these other quantities \cite{Cou08}. The dashed, dot-dashed and solid lines centered around $S_{0}=30.1 MeV$ in Fig.~\ref{fig5} represents $L$ values consistent with the analysis of $DR(Y(n)/Y(p))$ (Fig.~\ref{fig2}), $R_{i}(\alpha)$ and $R_{7}$ (Fig.~\ref{fig4}), respectively. The vertical line at $S_{0}= 31.6 MeV$ depicts the range of $L$ values obtained in ref. \cite{Bal08,Che05} from comparisons of IBUU04 calculations to the measured isospin diffusion data for $R_{i}(\alpha)$. Constraints from the isoscaling analyses in ref. \cite{She07} are not included as we have some concerns about their self-consistency, see ref. \cite{Sou08}.

\section{Context and Summary}

We have included, in Fig. ~\ref{fig5}, other recent constraints in the density dependence of the symmetry energy. The lower box centered at $S_{0}= 32 MeV$ depicts the range of $p_{0}$ values from analyses of Pygmy Dipole Resonance (PDR) data \cite{Kli07}. The upper box centered at $S_{0}= 32.5 MeV$ depicts the constraints reported in Ref. \cite{Dan08} from the analyses of nuclear surface symmetry energies. The Giant Monopole Resonance (GMR) data of ref. \cite{Tli07} does not directly provide $L$; however, ref. \cite{Bal08,Tli07} reports the GMR data to be consistent with the isospin diffusion analysis denoted by the line at $S_{0}= 31.6 MeV$.  The inset of Fig. ~\ref{fig5} shows the density dependence of the symmetry energy that results from combining our shaded region with the limiting values of $S_{0}= 30.2$ and $33.8 MeV$ provided by the PDR data \cite{Kli07}.

In summary, both isospin diffusion and double ratio data involving neutron and proton spectra have been consistently described by a QMD model. The analyses of all three observables provide consistent constraints on the density dependence of the symmetry energy. The results overlap with recent constraints obtained pygmy dipole and mass data. Some shifts in the boundaries of the constraints can be expected with improvements in the precision of the experimental data and in the understanding of theory. Nevertheless, the observed consistency between the different probes suggests that increasingly stringent constraints on the symmetry energy at sub-saturation density can be expected as the data improve and the model dependencies of their interpretations become better understood.

This work has been supported by the U.S. NSF via Grants PHY-0456903, 0555893, 0606007, 0800026, the MSU  High Performance Computing Center, the Chinese NSF via Grants 10675172, 10175093, 10235030, and the Chinese Major State Basic Research development program via contract No. 2

\end{document}